# Development of a 5 MHz frequency difference pre-multiplier for a short term frequency stability bench of the oscillators

Patrice Salzenstein, Xavier Jouvenceau, Xavier Vacheret, Gilles Martin and Franck Lardet-Vieudrin

FEMTO-ST Institute – Associated Laboratory to the LNE (Laboratoire National de métrologie et d'Essais)

CNRS UMR 6174, 32 avenue de l'Observatoire, F25044 Besançon Cedex

Besançon, France

patrice.salzenstein@lpmo.edu

*Abstract*—This paper reports the realization of a 5 MHz frequency difference x10 pre-multiplier, developed in the laboratory to replace an obsolete one. The principle we chose is to synthesize a 45 MHz and a 50 MHz from a reference signal and from the signal to be measured, and to subtract one to the other to generate a 5 MHz, whose precision on the measurement is increased by a factor 10. Obtained Allan variance $\sigma_y(\tau)$ at 1 s is $5.10^{-14}$ and output spectral density of phase noise floor is about -160 dBc/Hz at 5 MHz.

## I. Introduction

To measure the short term frequency stability of devices under test (DUT) such as Rubidium oscillators to be calibrated, we need to multiply the frequency difference between the frequency delivered by the DUT and the reference oscillator which is usually a quartz locked on a Cesium. The difference is too low to be measured directly by a frequency counter. We must keep additional noise of the frequency difference multiplier as low as possible in order not to degrade the performance of the DUT. A $10^5$ multiplication has to be realized at the end. The first stage of this multiplication is constituted by a new x10 frequency difference pre-multiplier. For next stages, we keep existing $10^4$ frequency difference multiplier for the first evaluation of the new modified calibration bench.

Bandpass of the developed pre-multiplier is about 100 kHz centered on 5 MHz. RF output power is 0 dBm. Attenuation of parasitic ray is lower than 60 dB. The main principle of the method is to generate a 50 MHz signal and mix it with a 45 MHz signal synthesized from the reference. 50MHz is derived from mixing the 8$^{th}$ with the 2$^{nd}$ harmonic of the DUT signal. For synthesis of the 45 MHz, the reference signal is combined with its 8$^{th}$ harmonic. The use of several x2 multipliers integrated on a board allows an optimal power rate associated with an excellent noise factor.

To measure its contribution to the noise, each board is realized twice. Output spectral density of phase noise is about -160 dBc/Hz at 5 MHz. For Fourier frequency between 0.1 and 100 Hz we obtain a -1/f slope corresponding a Flicker phase noise. The boards are then integrated together in a rack. Short term frequency stability is measured by rejecting a source. Allan deviation is better than $5.10^{-14}$ and $10^{-14}$ respectively at 1 and 10 s, comparable with existing system's performances [1-2].

The precise determination of the Allan deviation curve for integration time between 0.1 s and 100 s is one of the key parameter that laboratories need, to know the performances of the standard sources. One critical problem is the intrinsic noise of the system when one wants to evaluate the frequency short term stability from a measurement. This intrinsic noise may limit the noise floor that we want to measure. Commercial frequency difference multipliers don't allow to reach at the output of a bench, stabilities of few $10^{-14}$ at 5 MHz or 10 MHz. Pre-multiplier developed in the laboratory is close to state-of-the-art performances $\sigma_y(\tau=1s) = 5.10^{-14}$.

## II. Measurement Basics

The considered frequency domain goes from low frequencies to several hundreds of Mega Hertz. It is the most older explored domain, where techniques are well known, and where time and frequency analysis present good performances.

When oscillators deliver too close frequencies, very next, one to the other, sometimes, the difference between the two frequencies delivered by the oscillators is too small to be measured by classical methods. Let's recall that theses methods can be simply using a single frequency counter or performing the measurement using a beat note.

### III. PRINCIPLE OF THE FREQUENCY DIFFERENCE MULTIPLIER

A frequency difference multiplier with a factor of multiplication equal to N, allows to obtain a frequency $f_3$ from two close frequencies $f_1$ (DUT's frequency) and $f_2$ (reference frequency), that can be summarized using the following formula:

$$f_3 = f_2 + N.(f_2-f_1) = f_2 + N.\Delta f$$

The $f_3 - f_2$ difference is then measured by a counter. So, precision on the measurement of $f_1$ is increased by a N factor.

Different kind of frequency difference multipliers were manufactured in the past, but are not manufactured anymore, like the ADRET 4110A frequency difference multiplier using dividers. This one includes 4 stages that successively multiply by 10 the frequency difference, by mixing a frequency that is divided by 10 delivered by a Voltage Controlled Oscillator, with a frequency equal to $9/10.f_2$. The obtained signal is mixed with $f_3$. By cumulating four stages, the $10^4$ factor allows a 10 Hz frequency difference with a reference. Its intrinsic noise allows to measure difference of $10^{-12}$ during a 1s integration time.

Recently, a frequency difference multiplier was realized at 10 MHz and presents a $5.6.10^{-14} / \tau$ stability. [1]

We chose to design and realize a frequency difference multiplier with stages that consist in frequency multiplication. Principle is to multiply $f_1$ by 10 and $f_2$ by 9 and to mix them by subtracting and obtain $f_3$ that verifies:

$$f_3 = 10.f_1 - 9.f_2 = f_2 + 10.(f_2-f_1) = f_2 + 10.\Delta f$$

By using N stages in serial configuration, the $10^N$ frequency difference multiplication is realized as shown on the figure bellow, corresponding to N = 2.

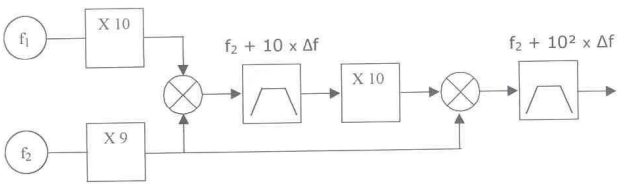

Figure 1: Schematics of a x100 frequency difference multiplier

### IV. CONCEPTION OF THE x10 FREQUENCY DIFFERENCE PRE-MULTIPLIER

As mentioned previously, the aim is to realize a pre-multiplier that can work in a 100 kHz band centered on 5 MHz. The output power level must be 0 dBm adapted on 50Ω. Attenuation of parasitic rays must be at least 60 dBc, that corresponds to an amplitude of the parasitic signal 1000 times less than the carrier.

The 5 MHz signal is multiplied by 10 or by 9 using frequencies multiplication based on hybrid junctions, associated with an image rejection mixer. The 50 MHz is derived from mixing the $8^{th}$ harmonic with the second harmonic of the DUT. Additional low-pass filters are inserted to reject harmonics, a band-pass filter is also placed at the output of each synthesis stage.

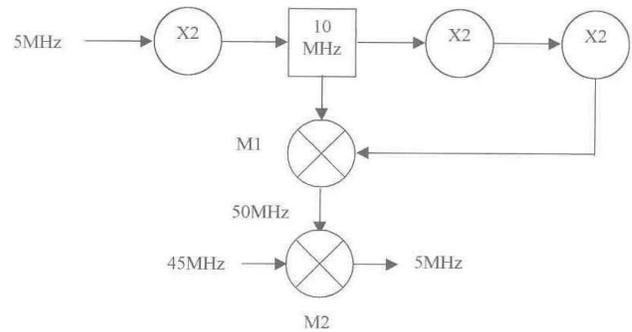

Figure 2: Schematic of the x10 frequency multiplier

To synthesize the 45 MHz signal, the reference signal is combined with its $8^{th}$ harmonic. Low pass and Band pass filters are also placed on the board like previously.

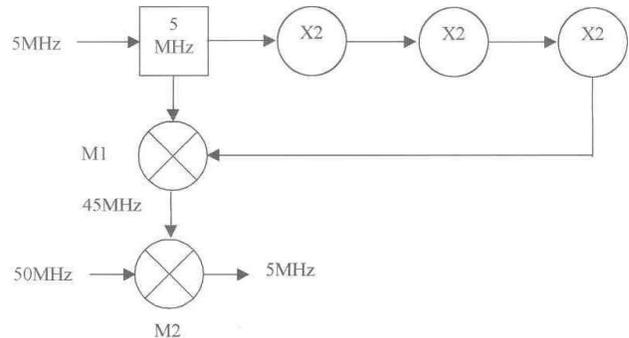

Figure 3: Schematics of the x9 frequency multiplier

The use of several x2 multipliers integrated on a board allows an optimal power rate, associated with an excellent noise factor. It takes advantage of the diodes bridge correction effect. For a 0 dBm 5 MHz input signal, the output levels at 5, 10, 15, 20, 25, 30 MHz are respectively -53, 0, -63,

-46, -87, -39 dBm. Parasitic rays at 20 MHz and 30 MHz are rejected after it through a low pass filter.

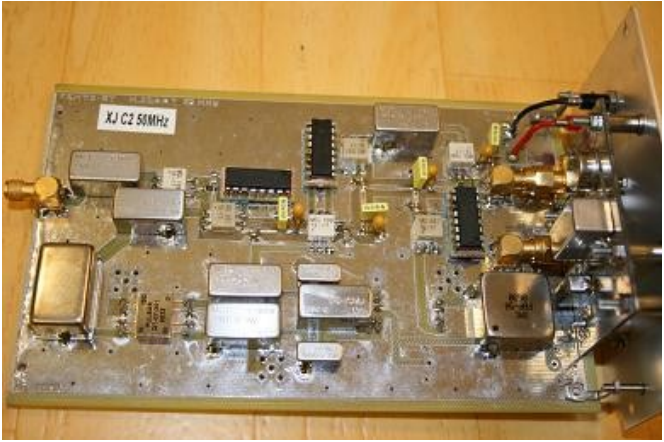

Figure 4: picture of the 50 MHz synthesis board

On the 50 MHz synthesis board, filter rejecting the 5 MHz and the 15 MHz signals after the first multiplication stage is placed in a metallic box to limit the undesirable electromagnetic coupling, close to the filtering selfs. It can be seen on the bottom left on the picture.

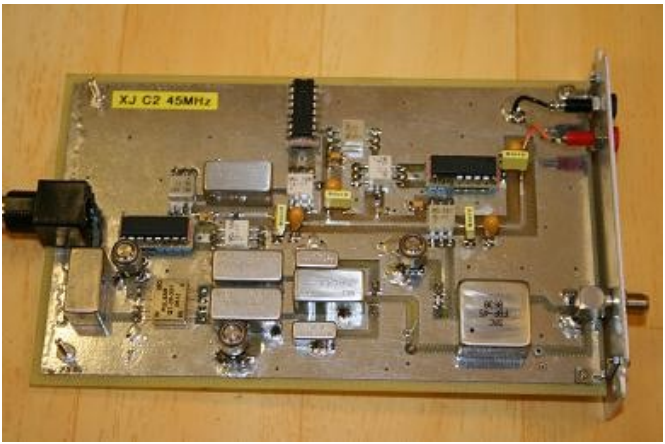

Figure 5: picture of the 45 MHz synthesis board

Each board was realized twice to allow them to be fully characterized by phase noise measurements. A 5 MHz signal delivered by a quartz is injected through both arms. One arm is phase shifted in quadrature. The signal is then sent to a fast Fourier Transform (FFT) analyzer.

The phase noise measure of two 50 MHz synthesis chain boards in parallel shows a -140 dB.rad²/Hz noise floor. It corresponds to -163 dBc/Hz brought back to the 5 MHz signal and with a 1/f flicker phase noise between 0.1 s and 100 Hz.

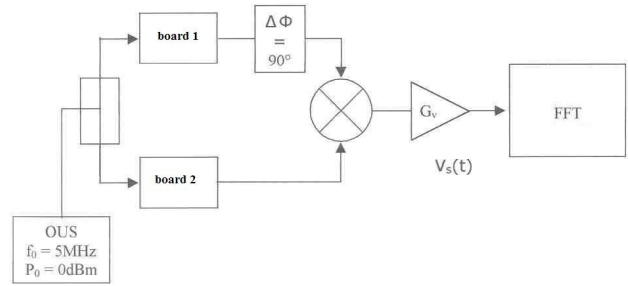

Figure 6: phase noise characterization principle

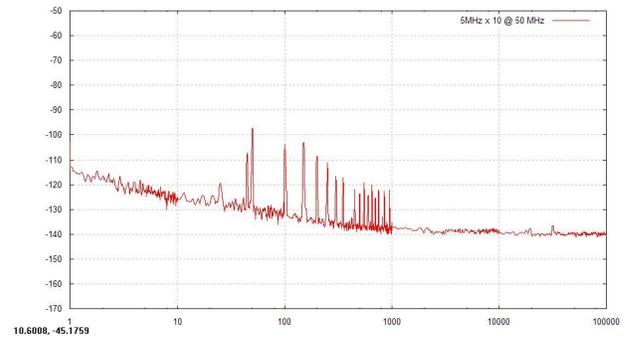

Figure 7: spectral density of phase noise measured on 50 MHz synthesis boards. Noise (dB.rad²/Hz) versus. Fourier frequency (Hz)

The phase noise measure of the 45 MHz synthesis chain boards in parallel shows a -137 dB.rad²/Hz noise floor. It corresponds to -159 dBc/Hz brought back to the 5 MHz signal and with a 1/f flicker phase noise between 0.1 s and 100 Hz.

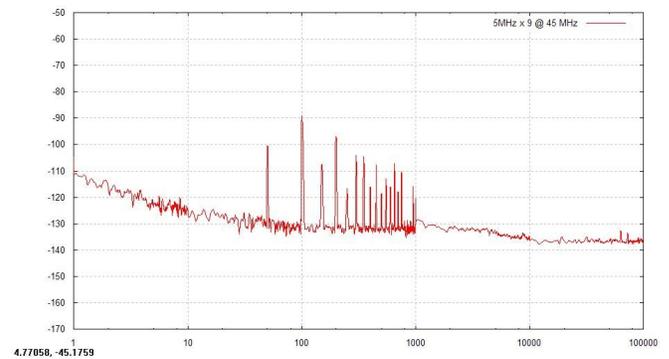

Figure 8: spectral density of phase noise measured on 45 MHz synthesis boards. Noise (dB.rad²/Hz) versus. Fourier frequency (Hz)

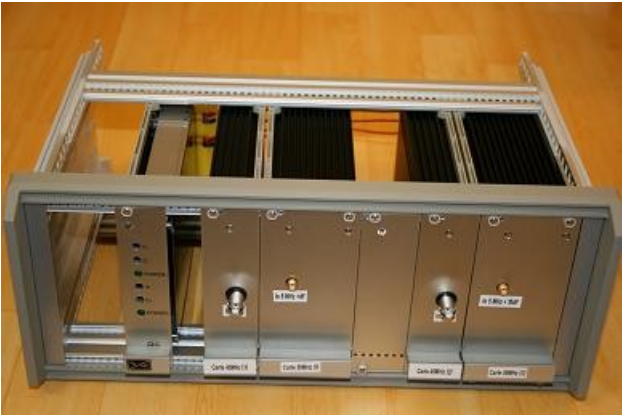

Figure 9: picture of the realized pre-multiplier

Noise floor is then measured in the time domain. The new x10 frequency difference pre-multiplier replaces the former one in a calibration bench. Obtained Allan variance $\sigma_y(\tau)$ at 0.1 s, 1 s, 10 s and 100 s is respectively $3.5 \cdot 10^{-13}$, $5 \cdot 10^{-14}$, $10^{-14}$ and $3.5 \cdot 10^{-15}$.

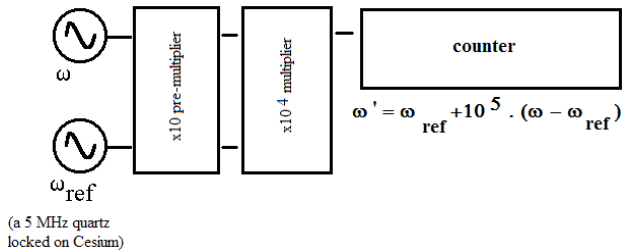

Figure 11: shematics of the x10 pre-multiplier is inserted in a calibration bench

## V. Conclusions and further work

The performances of the developed pre-multiplier enables it to be used in the calibration bench. The noise floor of this system is negligible enough to let us measure Rubidium and quartz oscillators. Allan variance $\sigma_y(\tau)$ at 1 s, $5 \cdot 10^{-14}$, is conform to the expected result.

To continue to improve the calibration bench, it remains to integrate the new distribution amplifiers developed in the laboratory [3]. At this time, the reference signal of the bench is a quartz locked on a signal delivered by a Cesium coming by an electric cable, from the Observatory of Besançon, but it should be soon possible to use a signal distributed trough an optical fiber local network between our laboratories [4].

## Acknowlegments

Authors would like to thank our colleague Jacques Groslambert, now retired, who began this project. We also thank the French national laboratory of metrology (LNE) that partly financed this project.